\documentclass{article}

\usepackage{arxiv}

\usepackage[utf8]{inputenc} % allow utf-8 input
\usepackage[T1]{fontenc}    % use 8-bit T1 fonts
\usepackage{hyperref}       % hyperlinks
\usepackage{url}            % simple URL typesetting
\usepackage{booktabs}       % professional-quality tables
\usepackage{amsfonts}       % blackboard math symbols
\usepackage{nicefrac}       % compact symbols for 1/2, etc.
\usepackage{microtype}      % microtypography
\usepackage{lipsum}

\usepackage{mathtools}
\DeclarePairedDelimiter\floor{\lfloor}{\rfloor}

\usepackage{cite}  % comment out for biblatex with backend=biber
\usepackage{graphicx,array}
\graphicspath{{figures/}{pictures/}{images/}{./}} % where to search for the images

\title{An Online and Nonuniform Timeslicing Method for Network Visualisation}

\author{ J. R. Ponciano$^{1}$, C. D. G. Linhares$^{1}$, E. R. Faria$^{1}$, and B. A. N. Traven\c{c}olo$^{1}$ 
        \\
        $^1$Faculty of Computing, Federal University of Uberl\^andia, Brazil\\
				\{jeanrobertop, claudiodgl, travencolo\}@gmail.com, elaine@ufu.br\\
       }

\begin{document}
\maketitle

\begin{abstract}
Visual analysis of temporal networks comprises an effective way to understand the network dynamics, facilitating the identification of patterns, anomalies, and other network properties, thus resulting in fast decision making. The amount of data in real-world networks, however, may result in a layout with high visual clutter due to edge overlapping. This is particularly relevant in the so-called \textit{streaming networks}, in which edges are continuously arriving (online) and in non-stationary distribution. All three network dimensions, namely \textit{node}, \textit{edge}, and \textit{time}, can be manipulated to reduce such clutter and improve readability. This paper presents an online and nonuniform timeslicing method, thus considering the underlying network structure and addressing streaming network analyses. We conducted experiments using two real-world networks to compare our method against uniform and nonuniform timeslicing strategies. The results show that our method automatically selects timeslices that effectively reduce visual clutter in periods with bursts of events. As a consequence, decision making based on the identification of global temporal patterns becomes faster and more reliable.
\end{abstract}

% keywords can be removed
\keywords{Temporal network visualization \and Nonuniform timeslicing \and Streaming network \and Network sampling.}

\section{Introduction}

Networks represent a useful and widely adopted structure to model systems from distinct areas, such as computer science, biology, sociology, and others~\cite{Estrada2015}. A network is defined in terms of nodes (instances) and edges (the relationship involving them)~\cite{ALB02a}. In this way, a network may be used to represent the World Wide Web (Web pages connected by hyperlinks), an organism cell (chemicals linked by chemical reactions), social interactions (any social relationship connecting individuals - e.g., friendship or collaboration), and many others~\cite{Estrada2015}. In several situations, using only information about nodes and edges may not be enough to represent and comprehend the relations in the network. In social network analysis, for example, the information of \textit{when} the connections occur may be crucial to describe such relations with less (or even without) loss of context. Networks in which the time information is considered, in addition to nodes and edges information, are studied in several disciplines and received different nomenclatures: temporal network, dynamic network, time-varying network~\cite{Holme2011}.

In temporal networks, the appearance of a new node or edge represents the occurrence of an event at the respective timestamp. In this type of network, all events are known and available to be used in the analysis (offline scenario)~\cite{Holme2011}. A temporal network has a delimited observation period, its data usually fit in primary memory and unrestricted random access is allowed. However, in several real-world applications, data are produced in a massive and continuous way (online scenario). These types of data are referred as \textit{data streams}. Examples include credit card transactions, phone calls, neuronal data, content sharing social networks, and others. In such applications, the volume of data may be so large that the storage may be impossible and mining tasks become more challenging~\cite{livroAggarwal}. Temporal networks can be used to model such streaming data~\cite{sampleStream1}. Since new events (edges) are continuously arriving in non-stationary distribution and typically at high speed, such networks are called \textit{streaming networks}~\cite{sampleStream1,Zhang2010,estrada2013communicability}. The non-stationary distribution of incoming data increases complexity and makes streaming network manipulation even more challenging.

Both temporal and streaming networks can be analysed by adopting different strategies. Statistical analysis represents a common approach and is useful to identify specific trends and patterns in the data, being used, for example, in connection prediction~\cite{estatisticaExemplo1} and burst analysis~\cite{estatisticaExemplo2}. When there is only a numeric output, however, it may represent a ``black-box'' to the user, thus impairing pattern comprehension. Another approach involves Information Visualisation~\cite{card1999readings,info_visualization}, whose strategies assist data analysis by providing interactive and graphical computational tools, thus including the user in the entire process of exploration and validation. An adequate Information Visualisation strategy allows a visual analysis that is as much intuitive as possible and also helps the user in finding unexpected patterns, anomalies, and other behaviours in the data, thus resulting in a fast and reliable decision making. Examples of recent visualisation methods applied to temporal and streaming networks can be found in ~\cite{cap,DyNetVis,Elzen2014,artigoSara,incremental2012,incrementalLayout,visualizacaoJoaoGama,dang2016timearcs,mtap}.

Visualisation techniques suitable for network analysis include node-link diagrams~\cite{SurveyStructural,CNO} and the \textit{Massive Sequence View (MSV)} layout~\cite{OriginalMSV,ElzenEnron}. Such layouts may suffer from visual clutter caused by the amount of information, impairing the analysis. To address this issue, all three network dimensions, namely \textit{node}, \textit{edge} (or \textit{connection}), and \textit{time}, can be manipulated to reduce clutter and improve readability. Node ordering techniques~\cite{Elzen2014,DyNetVis}, for instance, try to reduce edge overlapping by repositioning nodes, while sampling approaches~\cite{ResolutionLuis,EdgeSampling2018} select specific portions of the network to be analysed. Among these, the temporal sampling allows the selection of a subset of nodes and edges by reducing the observation time (e.g., considering only the first day of the network) or by timislicing the network, i.e., changing the network temporal resolution by  grouping edges from subsequent timestamps in a way that each timestamp may represent, for example, all edges from 1 minute or 1 day interval~\cite{ResolutionLuis,DyNetVis}).

The temporal resolution plays an important role in the layout construction and, consequently, in the visual analysis. In several scenarios, as, for example, when the networks are temporally sparse, an effective timeslicing may facilitate the analysis and highlight patterns that would be difficult to see using the original resolution~\cite{DyNetVis,Lee2019}. Choosing the length (in terms of number of timestamps) of each timeslice, however, is not a trivial task. Since it is usually chosen before the layout construction, the user needs to be a domain specialist, in order to know \textit{a priori} which timeslicing is the more adequate for the analysis given the expected event frequency distribution; otherwise it has to be empirically determined. A naive and widely used timeslicing approach is to consider timeslices of equal length to represent the network (uniform timeslicing)~\cite{ResolutionLuis,DyNetVis,EdgeSampling2018,CNO}. Despite the simplicity, the adopted value is a global, static, and pre-defined parameter that does not faithfully represent the number of events and their distribution. In both temporal and streaming networks, such distribution is non-stationary and changes over time, so the timeslicing method should consider this nonuniform behaviour. Although nonuniform timeslicing is often used in other contexts (e.g., multithreaded communication~\cite{MultithreadedCommunication}, data transfer~\cite{dataTransfer}, and systems performance degeneration analysis~\cite{performanceDegeneration}), it has only been considered recently in temporal network visualisation~\cite{wang}. Despite this recent advance, we are not aware of online and nonuniform timeslicing methods, necessary for streaming network analysis.

In this paper, we propose an online and nonuniform timeslicing method that automatically adapts the network temporal resolution scale according to the non-stationary distribution of events over time. Our method allows the identification of visual patterns, mostly global ones, that would be lost or difficult to find with a uniform timeslicing. Our focus is on online scenarios (streaming networks), which brings new challenges, as, for example, the need for fast (often real-time) methods and the immediate disposal of edges after processing~\cite{incrementalLayout}. Since any method developed for online scenarios can be applied in offline ones~\cite{sampleStream1}, our method can also be used to enhance temporal network analysis. Even knowing all events and having the possibility of unrestricted random access, improving the overall layout readability is already a challenging problem in temporal network visualisation~\cite{EdgeSampling2018,CNO}. Our proposal thus benefits both scenarios.

%The intuitive idea behind our approach is that if the network is having a lot of connections in a specific time interval, the resolution scale has to be decreased and each timestamp will then represent a bigger time interval. In this way, the visual clutter existent in this portion of the layout is reduced, favoring the identification of patterns.

The paper is organised as follows. Section~\ref{relatedWork} presents related concepts and discusses the related work. Section~\ref{method} describes the proposed timeslicing method. Section~\ref{experiments} presents two case studies using real-world networks. Section~\ref{limitations} describes the method's limitations. Section~\ref{conclusion} discusses conclusion and future work. %Section~\ref{limitations} describes the method limitations. Section~\ref{conclusion} discusses conclusion and future work.

%We conclude in Section~\ref{conclusion}.

%In the MSV layout, the vertical and horizontal axes represent the nodes and the timestamps, respectively. As it requires non-variant node positioning over time in order to preserve mental map, 

\section{Background and Related Work}
\label{relatedWork}

This section presents fundamental concepts and discusses strategies focused on network
visualisation. A discussion concerning
network dimensions (\textit{node}, \textit{edge}, and \textit{time}) manipulation is also provided.

\subsection{Temporal and Streaming Networks}

A temporal network can be represented by  $G = (V,E)$, where $V = \{n_{1}, n_{2}, ..., n_{N}\}$ is the set of nodes in the network and $E = \{e_{1}, e_{2}, ..., e_{M}\}$ is the set of edges. In our context, each edge $e_{i} = (n_{x}, n_{y}, t_{k})$ connects two nodes $n_{x}, n_{y} \in V$ at a particular and discrete timestamp $t_{k}$~\cite{cap,Holme2011}. Considering $t_{end}$ as the end of the observation period, $0 \leq t_{k} \leq t_{end}$. In fact, an edge that occurs at $t_{k}$ actually occurs in the interval $[t_{k},t_{k}+\tau)$, where $\tau$ is the temporal resolution~\cite{cap}. To simplify, self-edges (i.e., edges connecting a node to itself) are removed~\cite{cap}.

Streaming data that represent interactions among elements may be naturally represented as a \textit{streaming network}~\cite{sampleStream1,mcgregor2009graph}. In a telecommunication context, phone calls form a network between the involved phone numbers~\cite{Zhang2010}. In the same way, web pages and the links between them form a web network. Methods for processing streaming networks require efficient and real-time processing. This means that a streaming (or online) algorithm, besides the restricted access to the stream data (edges), must process the stream in a single scan, or in a small number of scans~\cite{Zhang2010}.

In temporal networks, all edges and nodes are known and available to be used in the analysis. In streaming, edges are continuously arriving, typically at high speed, and in a way that the volume of data does not fit in the primary memory~\cite{sampleStream1}. We define a streaming network $S = \{e_{1}, e_{2}, ..., e_{m}, ... \}$ as a temporal network $G = (V,E)$ as follows: $e_{i} = (n_{x}, n_{y}, t_{k})$, $e_{i} \in E$, represents an edge that occurs at a discrete timestamp $t_{k}$, $0 \leq t_{k} \leq \infty$, between $n_{x},n_{y} \in V$, $|V| \to \infty$. Note that it is possible to have more than one edge per timestamp. This definition is different from the ones that consider each edge arriving in a different timestamp~\cite{incrementalLayout,umaArestaPorTempo,etemadi2019pes}.

\subsection{Network Visualisation}

The employment of an effective temporal network visualisation strategy helps the user in the network evolution comprehension and facilitates the identification of patterns, anomalies, and other network properties. In this context, several visual strategies may be adopted, such as matrix-based~\cite{matrixcube,bach2014visualizing} and circular approaches~\cite{Elzen2014}, node-link diagrams~\cite{estrutural1,DyNetVis} and \textit{Massive Sequence View (MSV)} layouts~\cite{OriginalMSV,ElzenEnron}. Among these, node-link diagrams and MSV represent the best strategies when the task is to analyse the edge (event) distribution over time~\cite{Elzen2014,cap}. In this paper, we focus on MSV because of its visual scalability (considering the network size, in terms of number of nodes and edges) as well as its mental map preservation~\cite{CNO}.

%

%The node-link diagram (also known as \textit{Structural Layout}~\cite{DyNetVis}) is a classic network representation that spatially places nodes on the layout with edges connecting them~\cite{cap}. Each edge may contain a list of timestamps it appears in the network~\cite{Lee2019,cap}. The advantage of this representation is that it provides an overview of the network. The temporal dimension can be also explored through animation~\cite{surveyFoda}. The animation frames are updated at each timestamp, i.e., for each time, only the nodes and edges that are active at that moment are highlighted in the layout. Nodes and edges from the past timestamp(s) may also be shown with opacity to facilitate comprehension of the changes in the network structure. A major issue that affects this layout's readability is the choice of the animation transition speed~\cite{rey2010controlling,surveyFoda}. Nevertheless, because of the difficulty to maintain the mental map, strategies for reducing visual clutter in node-link diagrams (through node positioning \cite{circularStructural}, edge-bundling~\cite{edgebundling3}, and others) are generally ineffective even if combined with animation~\cite{cap}.

\begin{figure*}[ht!]
	\includegraphics[width=\linewidth]{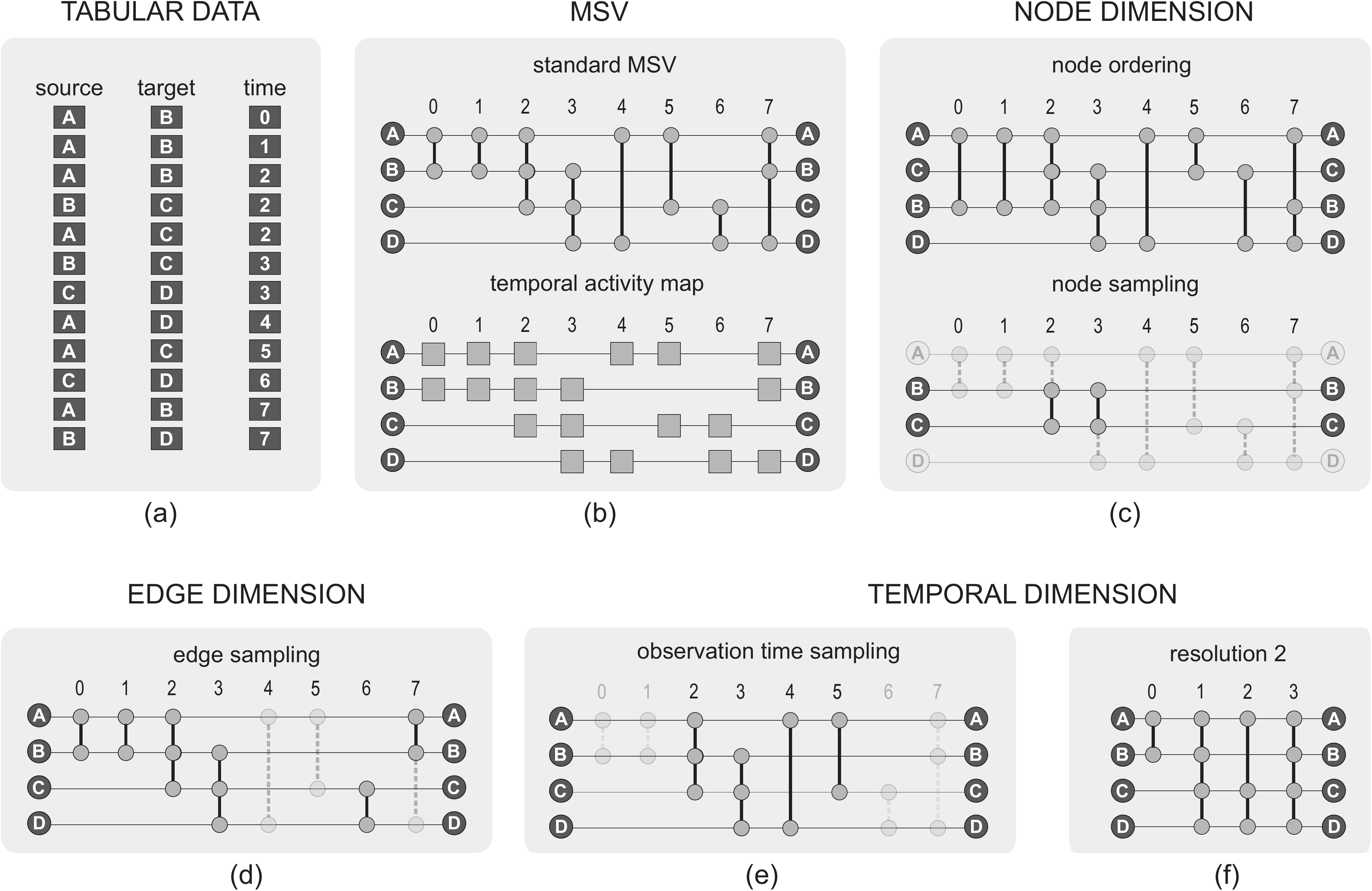}
	\centering
	\caption{MSV layout and possible types of manipulation. (a) Tabular (raw) data; (b) MSV layout: standard MSV, showing nodes and edges, and temporal activity map (TAM). In real-world networks that contain a large amount of data, MSV suffers from visual clutter. In this case, it is possible to improve the layout by manipulating the three network dimensions. (c) Node dimension manipulation -- ordering and sampling; (d) Edge dimension manipulation -- sampling; (e) Temporal dimension manipulation -- observation time sampling; (f) Temporal dimension manipulation -- timeslicing.}
	\label{fig:relatedwork}
\end{figure*}

The \textit{Massive Sequence View}~\cite{OriginalMSV,ElzenEnron} is a timeline-based layout~\cite{surveyFoda} similar to BioFabric~\cite{biofabric}. Its $x$-axis represents the timestamps and the $y$-axis represents the nodes of the network. In this layout, nodes cannot change their positions over time. Every time there is a connection (edge) between a pair of nodes, a vertical line is drawn linking them in the respective timestamp. The construction of the standard MSV layout using the tabular (raw) data is illustrated in Fig.~\ref{fig:relatedwork}(a,b). 

When applied in real-world networks with a large amount of data, node-link diagrams and MSV suffer from visual clutter caused by overlapping edges, and thus important patterns may not be perceived. The \textit{Temporal Activity Map (TAM)}~\cite{DyNetVis}, which omits all edges of the MSV layout and changes the shape of nodes from circles to squares for better sense of continuity, represents an alternative layout useful to identify patterns based only in the node activity (Fig.~\ref{fig:relatedwork}(b)). However, when the edge information is still needed or the amount of visual information in the layout is still large, these issues must be solved. %For this reason, several studies in literature propose methods to improve both layout readability and pattern identification~\cite{DyNetVis,ResolutionLuis,EdgeSampling2018,Elzen2014}. 

One strategy to enhance the layout is by changing the node positioning, which affects the length of the edges and, consequently, the number of overlapping edges and visible patterns. For this purpose, several node ordering algorithms have been proposed in the literature. For the MSV layout, examples include naive approaches, such as the ones based on the node appearance order, degree (in/out) and lexicographic, as well as more complex approaches, such as \textit{Optimized MSV}~\cite{ElzenEnron}, \textit{Recurrent Neighbors}~\cite{DyNetVis}, and \textit{Community-based Node Ordering (CNO)}~\cite{CNO}.  These strategies, however, are not suitable for streaming networks since they require all edges in primary memory. Figure~\ref{fig:relatedwork}(b,c) shows an example of node ordering and its impact on the edges.

Besides node positioning, sampling strategies can also improve layout readability by reducing the number of edges under analysis~\cite{CNO,EdgeSampling2018,EODShortPaper,joaoGamaParecido}. In this case, not only edges can be sampled, but also nodes -- only those edges connecting the sampled nodes are maintained --, as illustrated in Fig.~\ref{fig:relatedwork}(c-d).

\subsection{Network Timeslicing}
\label{relatedWorkResolucao}

Along with node positioning and sampling strategies, the temporal dimension can also be manipulated to improve layout readability. In this case, one possibility is to choose an observation time of interest (e.g., only the first or the second day of the network), as adopted by Zhao et al.~\cite{EdgeSampling2018} (Fig.~\ref{fig:relatedwork}(e)). Another possibility is to change the network temporal resolution scale through timeslicing strategies. Timeslicing in our context means that events from subsequent timestamps will be grouped in a single timestamp~\cite{ResolutionLuis}. The higher the temporal resolution scale, the more subsequent timestamps will be grouped into one and, as a consequence, the longer the time interval represented by the resulting timestamp.

In the simpler timeslicing approach (known as \textit{uniform timeslicing}), the temporal resolution scale is a global and static value, so all timestamps of the network represent the same length of time without considering cases in which the network has non-stationary event distribution. As an example, if each timestamp in resolution 1 represents a 20-second interval, then each timestamp in resolution 2 will represent a 40-second interval. Linhares et al.~\cite{DyNetVis} changes the timestamp in which each event occurs by following a uniform timeslicing as follows (Eq.~\ref{eq:resolucaoEstatica}).

\begin{equation}
\label{eq:resolucaoEstatica}
t_{\text{new}} = \left\lfloor \frac{ t_{\text{ori}} - t_{\text{s}}}{\tau} \right\rfloor \tau+t_{\text{s}}
\end{equation}

where $t_{\text{new}}$ is the new timestamp of the event, $t_{\text{ori}}$ is the timestamp of the event in the original temporal resolution, $t_{\text{s}}$ is the first timestamp of the network and $\tau$ is the desired resolution scale. Repeated events (edges) are considered as a single one if their timestamps are merged. As a result, one may identify temporal patterns that would be difficult to see in the original resolution, especially in temporally sparse networks~\cite{DyNetVis}. This timeslicing process is exemplified in Fig.~\ref{fig:relatedwork}(f), where in resolution 2 (new resolution defined by $\tau=2$) each pair of adjacent timestamps from the original network (Fig.~\ref{fig:relatedwork}(b)) are merged into one, thus the events represented in (A,B,0) and (A,B,1) become a single event in resolution 2 (A,B,0) and so on.

For convenience, several studies that visualise temporal networks employ uniform timeslicing in the networks under analysis~\cite{DyNetVis, EdgeSampling2018,cap,CNO,ResolutionLuis}. In~\cite{cap}, for instance, the the authors analyse a high-school network~\cite{highSchool} considering each timestamp of the MSV layout as a three-minute interval whilst the original network temporal resolution is a 20-second interval per timestamp. In the same way, the Enron network~\cite{enron2} was analysed using a specific temporal resolution in~\cite{EODShortPaper} and a different one in~\cite{DyNetVis}. Such global and static temporal resolution scale is empirically chosen through initial exploratory analyses or by a domain specialist that knows \textit{a priori} which one is adequate for the analysis given the event distribution. This scale does not consider the underlying network structure and thus may not faithfully represent the non-stationary distribution of events over time. Notwithstanding, in temporal networks all events are available in the visual analysis, so different scales may be tested until an adequate one is found.

%\red{adopt intervals with fixed duration that are defined by uniform timeslicing and} empirically chosen through initial exploratory analyses or by a domain specialist that knows \textit{a priori} which resolution scale is adequate for the analysis given the event distribution. Such global and static value \red{does not consider the underlying network structure and thus may not faithfully represent the non-stationary distribution of events over time. Notwithstanding, in temporal networks all events are available in the visual analysis, so different values for timeslice duration may be tested until an adequate one is found. }

In streaming scenarios, although one can adopt uniform timeslices as well, the employment of uniform approaches is even more difficult due to the non-stationary distribution of future events. Exploratory analysis may not be possible because usually there are no \textit{a priori} data to support the decision. Since the event distribution can change, considering only an initial set of events in the stream to support the choice may be inefficient as well. This is often ignored when it is assumed a uniform event distribution, as if the events came in consecutive timestamps~\cite{incrementalLayout,umaArestaPorTempo,etemadi2019pes}.

Wang et al.~\cite{wang} proposed a nonuniform timeslicing method for temporal network visualisation that creates timeslices with a balanced number of events (equal visual complexity) by using an approach similar to the histogram equalisation, well-established in the discipline of digital image processing. Their strategy (hereafter named \textit{Balanced Visual Complexity -- BVC}) uses more timestamps to represent high-activity periods (with bursts of events) and less timestamps otherwise. The method we propose in this paper goes in the opposite direction, i.e, we also consider that high-activity periods contain too much visual information, but we propose to represent them with higher resolution scales (consequently reducing the number of timestamps) instead of redistributing them in more timestamps. In the produced layout, the identification of global temporal patterns (e.g., birth and death of highly-active groups of nodes, bursts of events) is facilitated.  Moreover, contrary to BVC, our method runs online and thus is suitable for streaming network analysis. To the best of our knowledge, no other study has proposed online and nonuniform timeslicing methods for network visualisation.

\section{Online and nonuniform timeslicing method}
\label{method}

The idea behind our method is that intervals with the same length but that have different numbers of events must be represented by different resolution scales. Having more events leads to higher resolutions and thus in fewer timestamps, and so the amount of visual information is reduced to an appropriate level in an attempt to optimise the identification of patterns.

%\blue{Contrary to approaches that spend more bits/pixels to represent highly active time periods~\cite{timeslicing}, we reduce the amount of information to an appropriate level in an attempt to optimize the identification of patterns.}

Our method considers the number of events and their distribution on a fixed size window of timestamps (window of size $w_{size}$) to decide the temporal resolution scale that will be applied to the next window (i.e., next timeslice). Inside a window, the event distribution is considered by reducing the importance of older events using the forgetting mechanism fading sum, highly used in stream scenarios~\cite{livroJoaoGama,Gama2013}. %The fading sum is highly used in stream scenarios~\cite{livroJoaoGama}, in which the more recent the data is, the more relevant it is to the current state of the system.

Initially, we adopt the original resolution scale in the first window (cold start). From there, the resolution value for each subsequent non-overlapping window is calculated according to Eq.~\ref{newResolution}:

\begin{equation}
\sigma_{n} = \floor*{\delta . \sigma_{c} + (1 - \delta) . f_{s}(w_{size})}
\label{newResolution}
\end{equation}
where $\sigma_{n}$ is the resolution value for the next window, $\delta$ $(0 \leq \delta \leq 1)$ is a constant that determines the importance of the current resolution value ($\sigma_{c}$) in the computation of the new resolution, and $ f_{s}(w_{size})$ is the fading sum of all events in the current window, which is calculated according to a recursive formula (Eq.~\ref{aux1}):

\begin{equation}
	f_{s}(i) = \frac{x_{i}}{|T_{wc}|} + \alpha . f_{s}(i-1)
	\label{aux1}
\end{equation}
where $x_{i}$ is the number of events in position $i$ of the window, ${|T_{wc}|}$ is the number of timestamps considered by the window that presents at least one event, $f_{s}(1) = \frac{x_{1}}{|T_{wc}|}$ is the initial term, and $\alpha$ $(0 \ll \alpha \leq 1)$ is the fading factor. The higher the fading factor, the more importance is given to old events and, consequently, the higher the adopted resolution scale inside this window. If $\sigma_{n} = 0$, then $\sigma_{n}$ is set as the average value of all past resolutions, so large inactivity periods (i.e. without events) may be represented by a resolution scale different from the original. 

With the new resolution scale computed, it is possible to change the timestamp attribute of incoming events. Inspired by Eq.~\ref{eq:resolucaoEstatica}~\cite{DyNetVis}, we define the new timestamp $t_{new}$ of an event $e$ as:

\begin{equation}
t_{new}(e) = \floor*{\frac{t_{orig}(e) - t_{ini}}{\sigma_{n}}} + t_{ref}
\label{whereToPlot}
\end{equation}
where $t_{orig}(e)$ is the timestamp of $e$ in the original resolution, $t_{ini}$ is the first timestamp considered by the current window, $\sigma_{n}$ is the new resolution value (Eq.~\ref{newResolution}), and $t_{ref}$ is the timestamp that acts as a reference in order to apply the resolution scale in inactive timestamps. The value of $t_{ref}$ is computed when dealing with the first event of a new resolution (new timeslice) and is defined according to Eq.~\ref{aux2}:

\begin{equation}
	t_{ref} = \floor*{\frac{t_{ini} - t_{orig}(e')}{\sigma_{p}}} + t_{new}(e')
	\label{aux2}
\end{equation}
where $t_{orig}(e')$ is the original timestamp of the last event from the previous window (event $e'$), $\sigma_{p}$ is the previous resolution (the resolution applied on $e'$), and $t_{new}(e')$ is the timestamp of $e'$ in $\sigma_{p}$.

\begin{figure}[h]
	\includegraphics[width=0.9\linewidth]{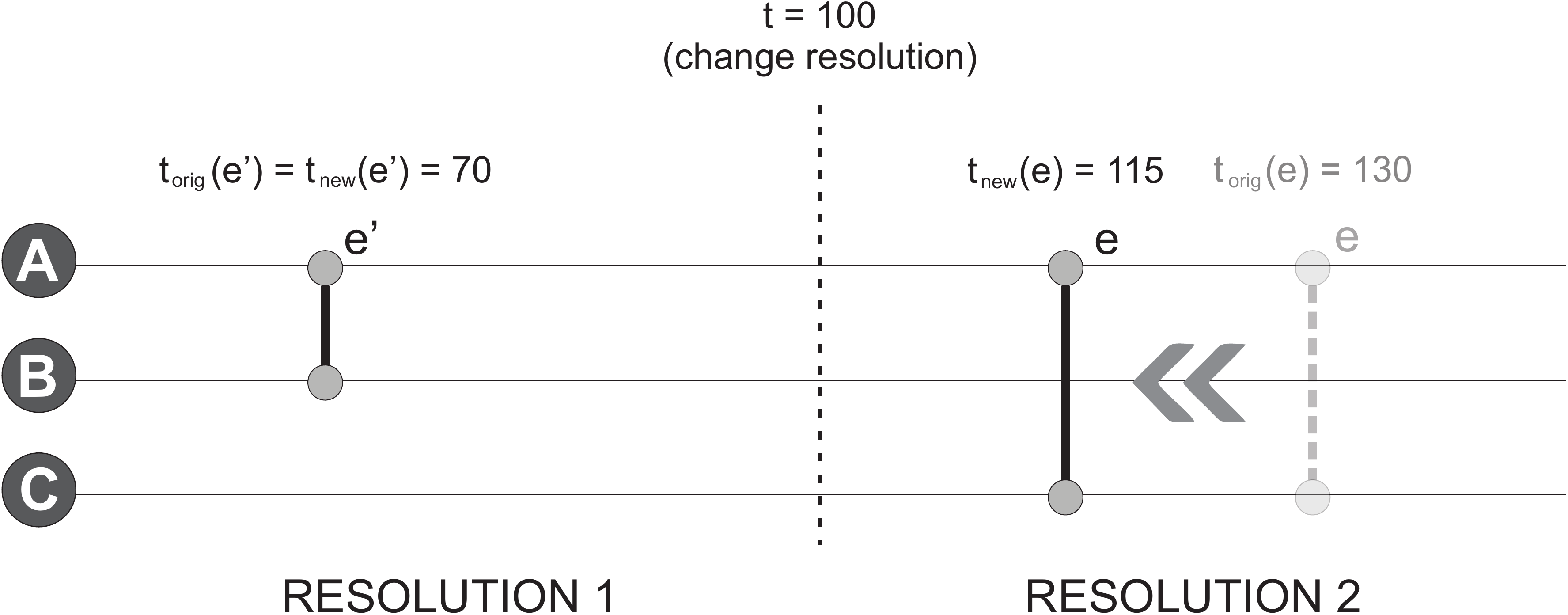}
	\centering
	\caption{Example of timeslicing using our proposal. In resolution 2, the timestamp of $e$ is changed from 130 to 115.}
	%	\caption{Example of a connection timestamp change due to the new resolution value. With the new resolution 2, the timestamp of $e$ is changed from 130 to 115. This happens because, with this resolution, each timestamp starting in $t = 100$ must represent the double of the time interval represented by a timestamp from resolution 1.}
	\label{fig:didaticaEq2}
\end{figure}

Figure~\ref{fig:didaticaEq2} shows an example of how the timestamp of an event is changed according to Eq.~\ref{whereToPlot}. In this figure, $t_{orig}(e') = t_{new}(e')$ because, up to this point, the original resolution (value 1) was maintained in the network. In $t = 100$, the new resolution value was computed (value 2) and so it was necessary to change the original timestamp of $e$. Each timestamp in resolution 2 is twice the time interval represented by a timestamp from resolution 1, thus $t_{orig}(e) = 130$ and $t_{new}(e) = 115$. Assuming an event $x$ with $t_{orig}(x) = 131$, then $t_{new}(x)$ would be equal to 115 as well, and so on. As stated, Eq.~\ref{whereToPlot} takes into account inactivity periods, respecting their occurrence in the converted timestamps.

\section{Case Studies}
\label{experiments}

\begin{figure*}[!t]
	\includegraphics[width=\linewidth]{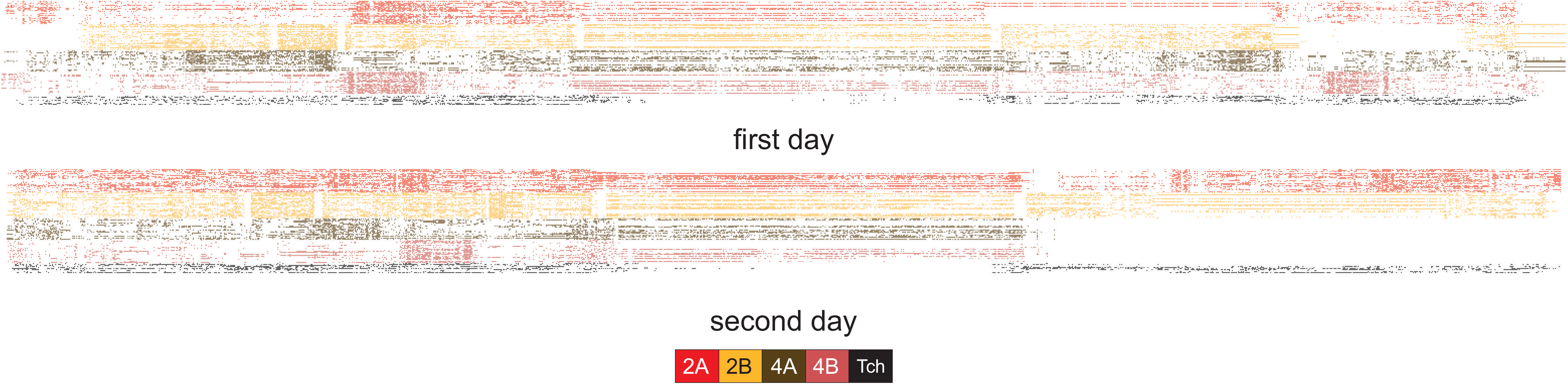}
	\centering
	\caption{TAM layout showing four classes and all teachers of the Primary School network using resolution 1 (original). The interval between both days (from 5.21pm to 8.29am) does not present any edge and was omitted due to its size in the layout. Nodes are grouped according to the classes and grades. The ``Tch'' profile refers to the teachers of the school. The layout is horizontally large, dense and has few visible patterns, as, for example, the absence of classes 4A and 4B near the end of the second day.}% Figure~1 of the supplementary material shows the same image in a better quality and with the omitted interval.}
	\label{fig:res1day1primaryschool}
\end{figure*}

In this section, we present visual analyses of two real-world temporal networks manipulated timestamp-by-timestamp to simulate streaming scenarios. Our goal is to compare our nonuniform timeslicing method against the original network resolution, uniform timeslicing approaches, and BVC~\cite{wang}. In all analyses, we consider resolution 1 (Res. 1) as the original resolution of the network and $\delta = 0.2$ as the importance of the current resolution in the computation of the new one (see Eq.~\ref{newResolution}). To validate our method and illustrate its application, we rely on MSV~\cite{ElzenEnron} and TAM~\cite{DyNetVis}. All experiments were performed using the software DyNetVis~\cite{DyNetVis}, which implements our method and all layouts and features presented in this section. DyNetVis is freely available at \url{www.dynetvis.com}.

%and weight equals 0.2 to the current resolution value in the computation of the new one (see Equation~\ref{newResolution}).

\subsection{Primary School}

The first network, \textit{Primary School}~\cite{primarySchool1,primarySchool2}, represents contacts involving teachers and students of a primary school between October $1^{st}$ - $2^{nd}$ of 2009. This network contains 242 nodes and 125,773 edges distributed in 5,846 timestamps. The original temporal resolution is 20 seconds, which means that each timestamp in Res. 1 comprises a 20-second interval. The data represent contacts from the first to fifth grade, each of them having two classes (A and B). In this network, the majority of contacts occurs between students of the same class and each class has an assigned teacher~\cite{primarySchool2}.

Figure~\ref{fig:res1day1primaryschool} presents the TAM layout for four classes and all teachers of the Primary School network in resolution~1 (original). The nodes are grouped according to the classes and grades. The layout is horizontally large (due to the number of timestamps), which impairs the identification of global patterns and requires more screen space and scrolling, which impairs the user's perception of temporal changes during the network evolution (mental map preservation~\cite{mentalmapSugeridoCGF}). Moreover, the layout is dense (a lot of events over time) and only a few patterns are easily identified, as, for example, the absence of classes 4A and 4B students near the end of the second day. The network does not register contacts during sports activities~\cite{primarySchool2}, so it is possible to assume that these classes were involved in such activities or dismissed. Another possibility is that the students were taking exams or other activities without interacting with each other.  

%Since the exhibition of edges in the layout is directly related to the visual clutter due to edge overlapping and

%To improve pattern identification, we applied our adaptive resolution in the Primary School network. Figure~\ref{fig:graficosPrimarySchool} depicts how the Fading Factor ($FF$) and the sliding window size ($w_{size}$) affect the resolution and, consequently, the layout and the perception of patterns. In the figure, the whole network is considered (activity in day 1, interval, activity in day 2).By comparing the plots vertically, in which the $FF$ value is the same, it is possible to see that a large window makes that the perception of changes in the level of node activity be late, delaying the resolution change. 

\begin{figure}[ht!]
	\includegraphics[width=0.5\linewidth]{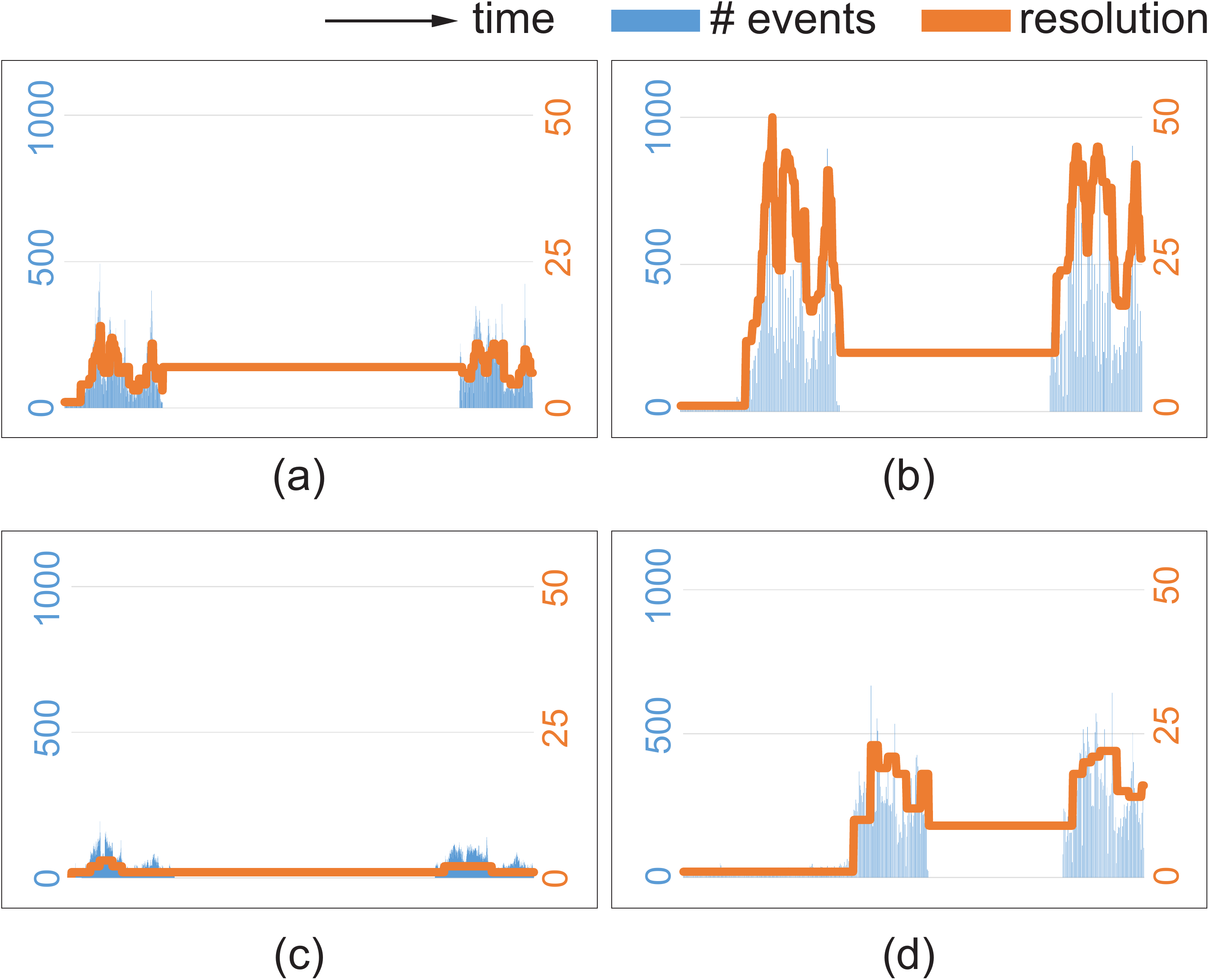}
	\centering
	\caption{Our nonuniform timeslicing and the relation between the adopted resolution scales and the event distribution for the Primary School network. (a) $w_{size}=50$ and $FF = 0.9$ (1,443 timestamps). (b) $w_{size}=50$ and $FF = 0.99$ (353 timestamps). (c) $w_{size}=200$ and $FF = 0.9$ (4,880 timestamps). (d) $w_{size}=200$ and $FF = 0.99$ (541 timestamps). The choice of the Fading Factor ($FF$) and the window size ($w_{size}$) affects the resolution scale and, consequently, the layout and visible patterns.}
	\label{fig:graficosPrimarySchool}
\end{figure}

To improve pattern identification, we applied our nonuniform timeslicing method in the Primary School network. Different values for Fading Factor ($FF$) and window size ($w_{size}$) were evaluated and their impact in the adopted resolution scales is shown in Fig.~\ref{fig:graficosPrimarySchool}. In the figure, the whole network is considered (activity in day 1, interval, activity in day 2).
By comparing the plots in which the $FF$ value is the same ($FF = 0.9$ in (a,c) and $FF = 0.99$ in (b,d)), it is possible to see that a large window makes that the perception of changes in the number of events be late, delaying the resolution change. As a consequence, patterns related to these changes may be lost or identified only many timestamps later. This is especially relevant in streaming network analysis, in which the past data may have already been discarded. By comparing the plots in which the $w_{size}$ is the same ($w_{size} = 50$ in (a,b) and $w_{size}= 200$ in (c,d)), one can notice higher resolution values when adopting higher $FF$. This is expected since high $FF$ values increases the importance of old events. Finally, the plots show the resolution adopted in the interval between both days of the network, in which there is no event. This value is computed based on the average value of the past resolutions. This decision is related to the space of the layout required to represent such interval, which would be many times greater in the original resolution. 

\begin{figure}[h]
	\includegraphics[width=\linewidth]{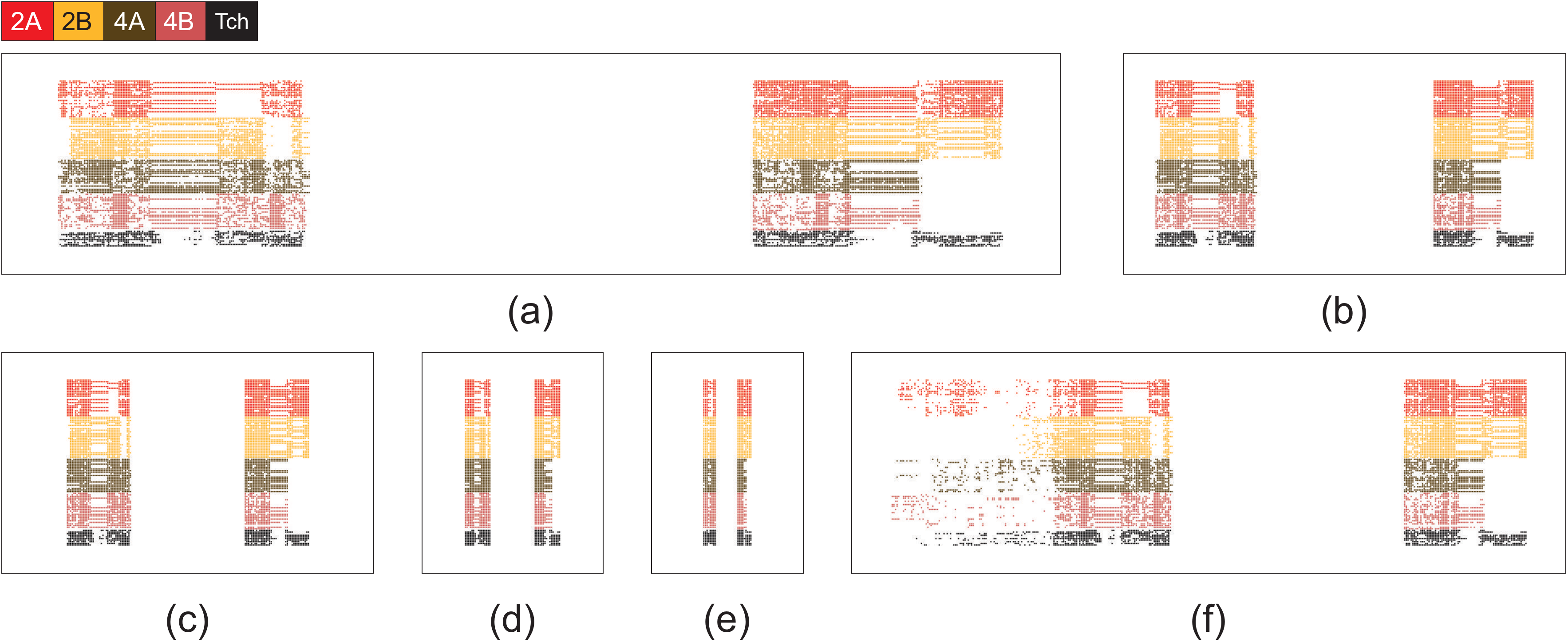}
	\centering
	\caption{TAM layouts showing four classes and all teachers of the Primary School network according to our nonuniform timeslicing method and five uniform resolution scales. (a) Res. 10. (b) Res. 25. (c) Res. 39. (d) Res. 100. (e) Res. 200. (f) Our method ($w_{size} = 100$ and $FF = 0.99$).}
	\label{fig:resolutions_school}
\end{figure}

Figure~\ref{fig:resolutions_school} shows TAM layouts for different timeslicing scales considering the same four classes and teachers from Fig.~\ref{fig:res1day1primaryschool}. Figure~\ref{fig:resolutions_school}(a-e) shows TAM layouts for uniform timeslices adopting different resolutions (Res. 10, 25, 39, 100, and 200, respectively). Figure~\ref{fig:resolutions_school}(f) shows the TAM layout generated by our nonuniform timeslicing method ($w_{size} = 100$ and $FF = 0.99$, chosen empirically). Resolutions 10, 25, and 39 (Fig.~\ref{fig:resolutions_school}(a-c)) were chosen because they represent the lower, the average and the higher resolution values adopted by our method for this network. Resolutions 100 and 200 (Fig.~\ref{fig:resolutions_school}(d-e)) are arbitrary values. As expected, higher resolution values generate denser and (horizontally) smaller layouts, which impairs the visual analysis and the identification of patterns. Our method (Fig.~\ref{fig:resolutions_school}(f)), however, automatically defines resolution scales that represent appropriate levels of visual density.

\begin{figure*}[!t]
	\includegraphics[width=\linewidth]{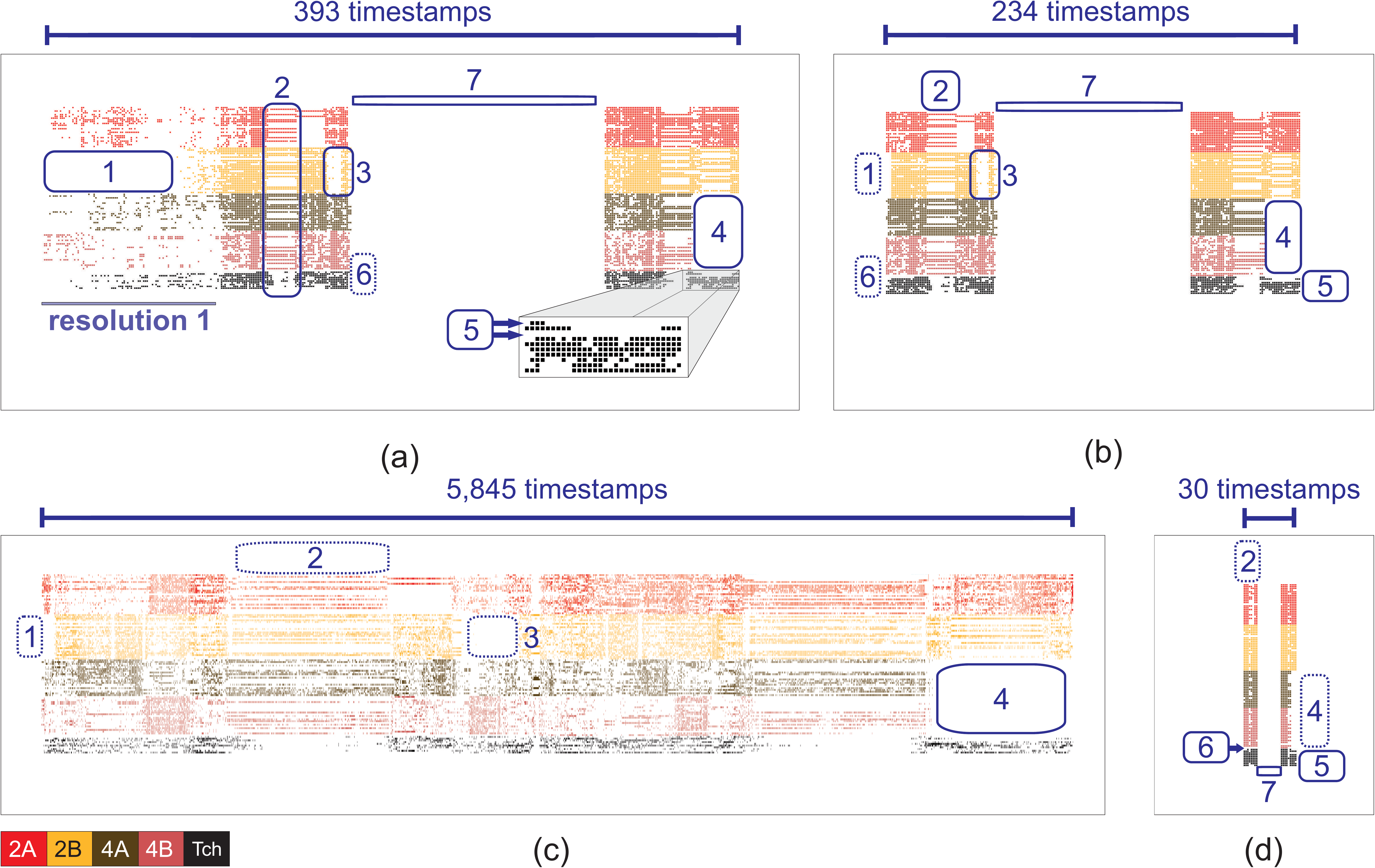}
	\centering
	\caption{Visible patterns in TAM layouts generated by different timeslicing approaches for the Primary School network. (a) Our method ($w_{size} = 100$ and $FF = 0.99$). (b) Res. 25. (c) BVC. (d) Res. 200. A maximum of seven patterns can be identified: (1) class 2B students joined the network after the others; (2) lunch break; (3) no interaction involving class 2B students near the end of the $1^{st}$ day; (4) absence of classes 4A and 4B near the end of the $2^{nd}$ day; (5) two teachers left the network after lunch in the $2^{nd}$ day; (6) some students did not join the network in the $1^{st}$ day; and (7) inactivity period. Continuous rectangles represent patterns considered as easy to identify. Dotted rectangles represent patterns with difficult perception.} %Figure~2 of the supplementary material shows BVC's layout in a better quality.}
	\label{fig:comparaPadroesEscola}
\end{figure*}

The adopted timeslicing strategy highly affects pattern identification. Figure~\ref{fig:comparaPadroesEscola} presents visual analyses over the TAM layouts generated by our method (adopting $w_{size} = 100$ and $FF = 0.99$, Fig.~\ref{fig:comparaPadroesEscola}(a)) and by uniform timeslices using resolutions~25~and~200 (Fig.~\ref{fig:comparaPadroesEscola}(b,d), respectively). These are the same layouts from Fig.~\ref{fig:resolutions_school}(f,b,e). The layout generated by BVC is also considered (Fig.~\ref{fig:comparaPadroesEscola}(c)). % -- for a high quality image of BVC's layout, please refer to Fig.~2 at the supplementary material). 
In the best-case scenario, at least seven patterns can be identified: (1) all students from class 2B joined the network after the other classes and the group of teachers; (2) lunch break -- several students go home for lunch, which reduces the number of interactions in such interval~\cite{primarySchool2}; (3) there is no interaction involving class 2B students in a time interval near the end of the first day -- probably due to sports activities~\cite{primarySchool2}; (4) absence of classes 4A and 4B students near the end of the second day; (5) two teachers left the network after lunch in the second day -- probably the teachers from classes 4A and 4B; (6) there are students that did not join the network in the first day; and (7) inactivity period due to the absence of classes (from 5.21pm to 8.29am).

Our method allows the identification of all seven patterns (Fig.~\ref{fig:comparaPadroesEscola}(a)), being six of them considered as easy to found (1-5,~7). Pattern~6 is harder to identify because of the method's cold start (adoption of resolution 1 at the beginning of the layout), which pollutes the layout and impairs the perception of this pattern. Although this original resolution serves only as a start point, considering it inside the layout facilitated the perception of pattern~1. This pattern can also be noticed when adopting only resolution~1 in the analysis (see Fig.~\ref{fig:res1day1primaryschool}), but not as fast as with our proposal. Patterns 2,~5, and~6, on the other hand, cannot be identified with resolution 1 (Fig.~\ref{fig:res1day1primaryschool}).
By adopting a uniform timeslicing using resolution~25 (Fig.~\ref{fig:comparaPadroesEscola}(b)), all seven patterns can be identified as well, five of them being considered as easy to found (2-5,~7) and two of them being a little harder (1,6). Although this layout allows the identification of all patterns, recall that this resolution is the average value considered by our nonuniform method, which supports our method's quality. By considering BVC in the network analysis, one may see the event redistribution caused by BVC's histogram equalisation. As a consequence, pattern~7 is lost. Due to the number of timestamps, patterns~5-6 are also lost and patterns~1-3 are difficult to perceive. Only pattern~4 is considered easy to found. Such pattern, however, is also easy to identify with our method and with uniform resolutions~1 and~25. By using the uniform resolution 200 (Fig.~\ref{fig:comparaPadroesEscola}(c)), patterns~1 and~3 are lost and only patterns~5-7 are considered as easy to identify. Note that pattern 6 is more easily perceived in this layout, so a uniform timeslicing that considers a higher resolution value may be useful in specific scenarios as well. 

In summary, our method reduces the amount of visual information to an appropriate level that optimises the identification of global patterns that are lost or difficult to perceive with other timeslicing strategies, including BVC. Our layout for this network spent 393 timestamps (against 5,845 from BVC) while preserving all seven analysed patterns. Less timestamps leads to less screen space and decreases the need of (horizontal) scrolling, which tends to facilitate the perception of temporal changes in the network (better mental map preservation). Considering uniform timeslicing, one should test different resolution scales until the better one is found. This approach, however, is only possible when dealing with (non-streaming) temporal networks (see Section~\ref{relatedWorkResolucao}). Our method not only provides adequate timeslices, but is suitable for streaming scenarios in which events are continuously arriving in non-stationary distribution.

\begin{figure}[!h]
	\includegraphics[width=0.5\linewidth]{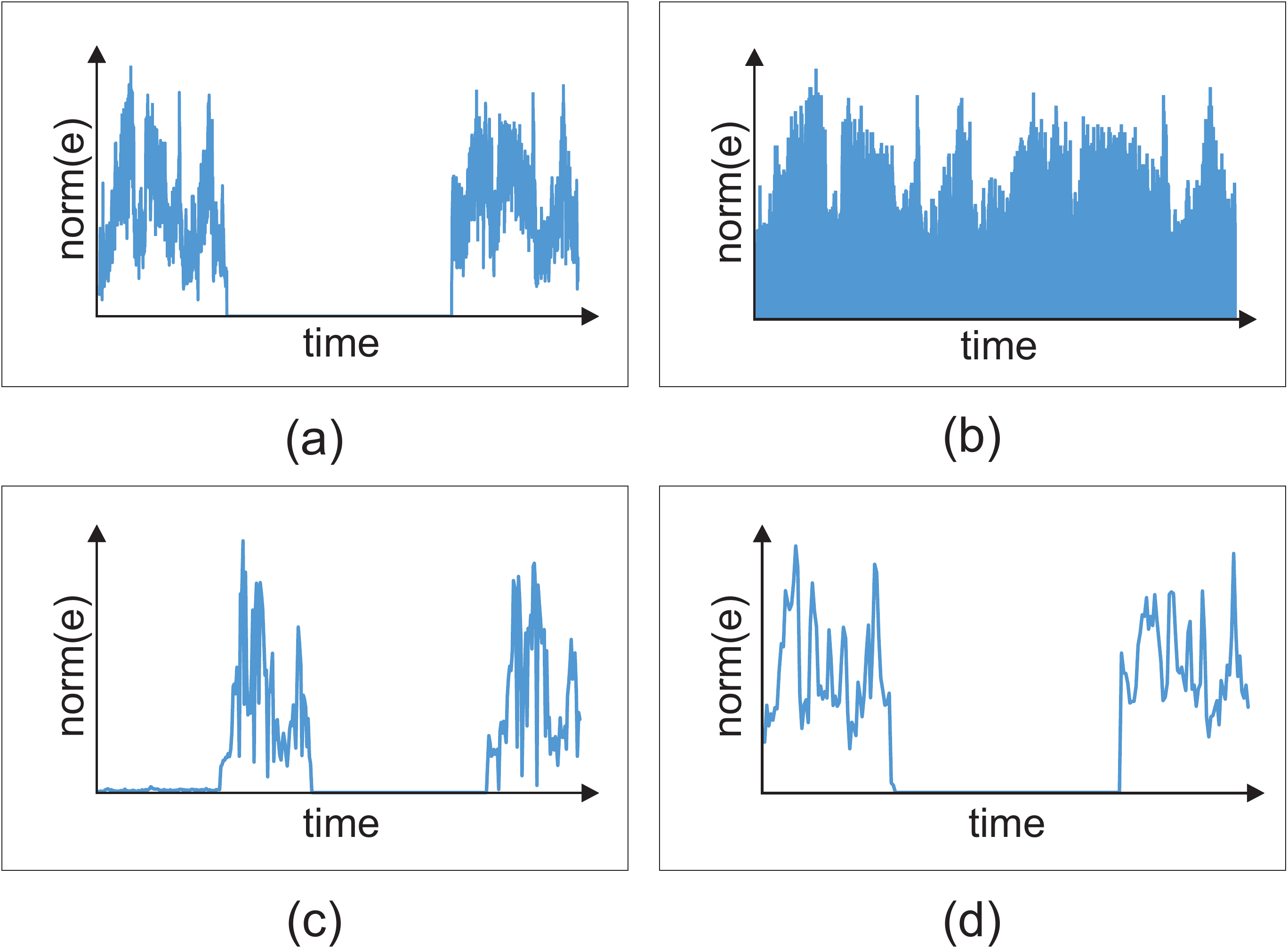}
	\centering
	\caption{Spread of events according to different timeslicing approaches for the Primary School network. (a) Res.~1. (b) BVC. (c) Our method ($w_{size} = 100$ and $FF = 0.99$). (d) Res.~25. ``norm(e)'' refers to the normalisation of the number of events to values between 0~and~1.}
	
%	Jeitao dos graficos. (a) Res 1 (5,846 timestamps). (b) Wang BVC - Balanced Visual Complexity) (5,846 timestamps). (c) Nosso w100,f0,99) (393 timestamps). (d) Res 25 (234 timestamps). norm(e) = funcao que normaliza o edge distribution entre [0,1].}
	\label{fig:graficosEscolaJeitao1}
\end{figure}

Figure~\ref{fig:graficosEscolaJeitao1} shows the spread of events over time according to different timeslicing approaches: the original resolution, BVC, our method ($w_{size} = 100$ and $FF = 0.99$), and uniform Res. 25. The absence of events in the middle of plots (a,c,d) corresponds to the inactivity period between both days of the network. While BVC (Fig.~\ref{fig:graficosEscolaJeitao1}(b)) changes the event distribution because of its histogram equalisation -- which impacts pattern identification --, our method (Fig.~\ref{fig:graficosEscolaJeitao1}(c)) provides a distribution similar to those from uniform approaches (Fig.~\ref{fig:graficosEscolaJeitao1}(a,d)). Since our timeslicing adopts the original resolution in the first window (cold start), one may see a ``shift'' in the time dimension at the plot (Fig.~\ref{fig:graficosEscolaJeitao1}(c)).

\begin{figure}[ht]
	\includegraphics[width=0.5\linewidth]{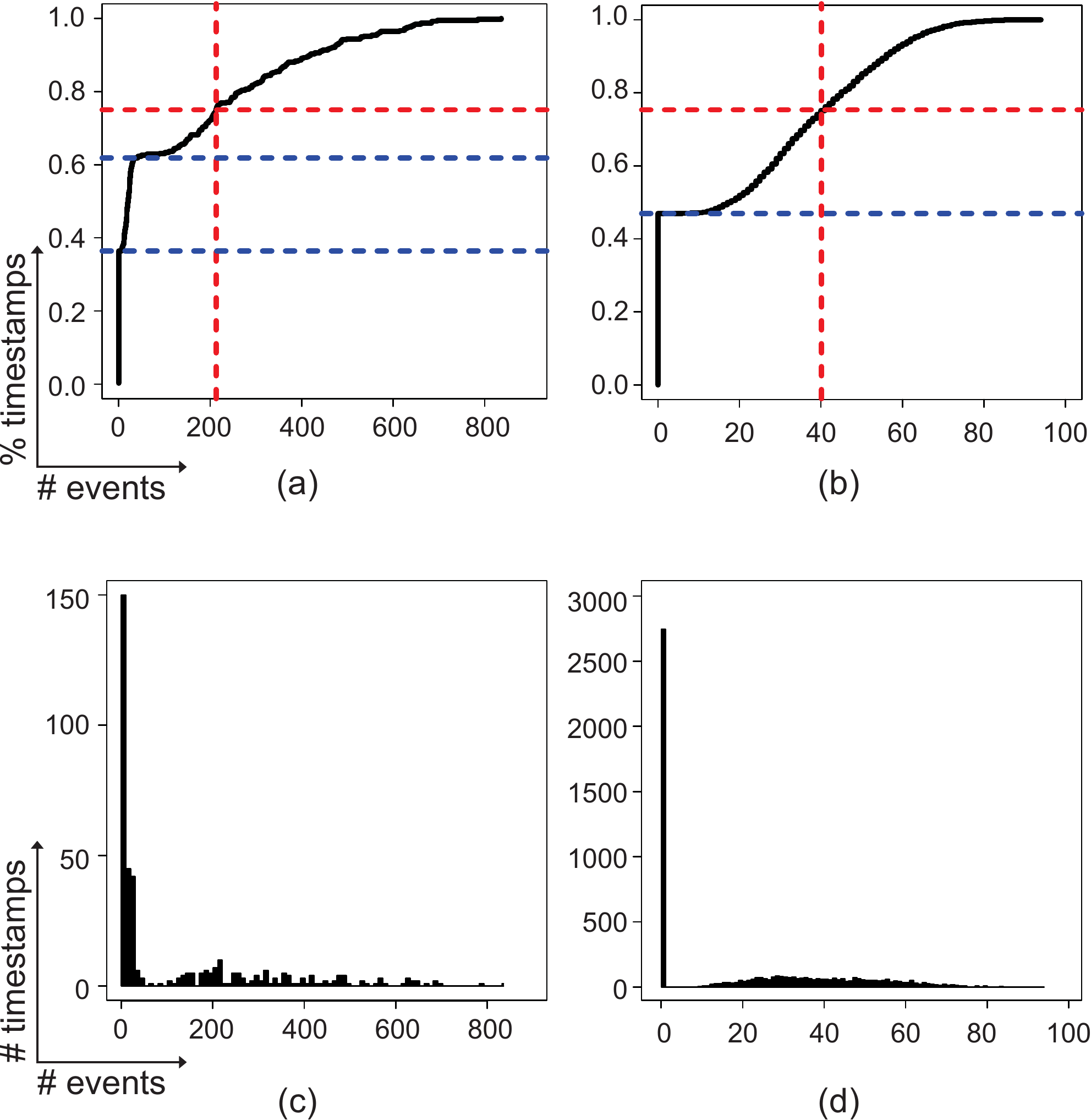}
	\centering
	\caption{Empirical cumulative distribution function (ECDF) and event distribution (ED) considering the events from the Primary School network. (a) ECDF our method ($w_{size} = 100$ and $FF = 0.99$). (b) ECDF Res.~1. (c) ED our method (393 timestamps, $w_{size} = 100$ and $FF~=~0.99$). (d) ED Res.~1 (5,846 timestamps).}
	\label{fig:ecdf_escola}
\end{figure}

Figure~\ref{fig:ecdf_escola} shows the empirical cumulative distribution function (ECDF) considering the events from our method's layout ($w_{size} = 100$ and $FF = 0.99$, Fig.~\ref{fig:ecdf_escola}(a)) and from resolution 1's layout (Fig.~\ref{fig:ecdf_escola}(b)). Our method produces less timestamps without events when compared with the original resolution (36.6\% vs 47\% -- blue dotted lines), which is justified by the resolution scale used in inactivity periods, that is different from the original one. Furthermore, 25.4\% of the time contains very few events in our layout (cold start window). By observing the third quartile (red dotted lines), after 75\% of the time our layout contains a maximum of 213 events per timestamp (26\% of the maximum number of events per timestamp), while in resolution 1 the number of events per timestamp is almost 43\% of the maximum number of events per timestamp (40 out of 94 events).

\begin{figure}[ht]
	\includegraphics[width=0.75\linewidth]{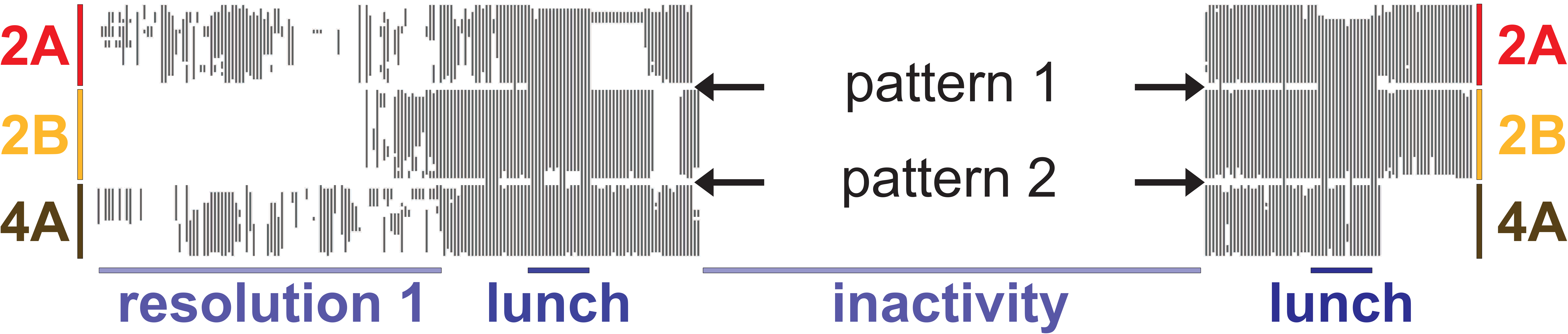}
	\centering
	\caption{MSV layout with our nonuniform timeslicing ($w_{size} = 100$ and $FF = 0.99$) showing the interactions among classes 2A, 2B and 4A. Pattern~1: many interactions between 2A and 2B during lunch. Pattern~2: few interactions between 2B and 4A in the network.}

	%MSV layout with our nonuniform timeslicing ($w_{size} = 100$ and $FF = 0.99$) showing the interactions among classes 2A, 2B and 4A. This layout reaffirms some of the patterns previously described and allows the identification of new ones: many interactions between 2A and 2B during lunch (pattern~1); few interactions between 2B and 4A in the network (pattern~2).}
	\label{fig:ArestasPrimarySchool}
\end{figure}

The visual analysis can be performed from a different perspective by showing only edges, as illustrated in Fig.~\ref{fig:ArestasPrimarySchool}, that shows the interactions involving classes 2A, 2B, and 4A over a MSV layout generated by our nonuniform timeslicing ($w_{size} = 100$ and $FF = 0.99$). This layout reaffirms: (i) students from class 2B joined the network after the others; (ii) students from class 4A left the network earlier than the others in the second day; (iii) the absence of the majority of 2A students, as well as 2B students, during a period after the lunch break in the first day. Besides, this layout reveals new patterns, such as the perception that the only two students from class 2A that stayed in the network during the time interval after lunch in the first day connected to one another. Moreover, the layout shows that students from one class have few interactions with students from other classes, with the majority of these interactions occurring during lunch. Not least, students from the $2^{nd}$ grade interact more between themselves than with class 4A. This behaviour is also observed in the rest of the network (a lot of interactions among students of the same grade and few interactions involving different grades). These situations are expected in the network~\cite{primarySchool2} and easily perceived in this layout.

\subsection{Enron}

The second network, \textit{Enron}~\cite{enron,enron2}, contains email communications from Enron Inc., a former energy company involved in the biggest American accounting fraud~\cite{graphScope}. The network is composed of 148 nodes and 24,667 edges distributed in 1,346 timestamps~\cite{DyNetVis}. Its original temporal resolution is 1 day. Enron was studied by several works in literature, from offline visual analysis using MSV (e.g.,~\cite{DyNetVis,ElzenEnron,EdgeSampling2018}) to streaming-fashion mining tasks (e.g.,~\cite{graphScope}). Unlike the Primary School network, whose number of events varies a lot in each day and which contains a large time interval without any event (the period between the two days), the Enron network presents a growing number of events over time. We applied our method in the network to analyse the evolution of the resolution under this circumstance.

\begin{figure}[ht!]
	\includegraphics[width=0.5\linewidth]{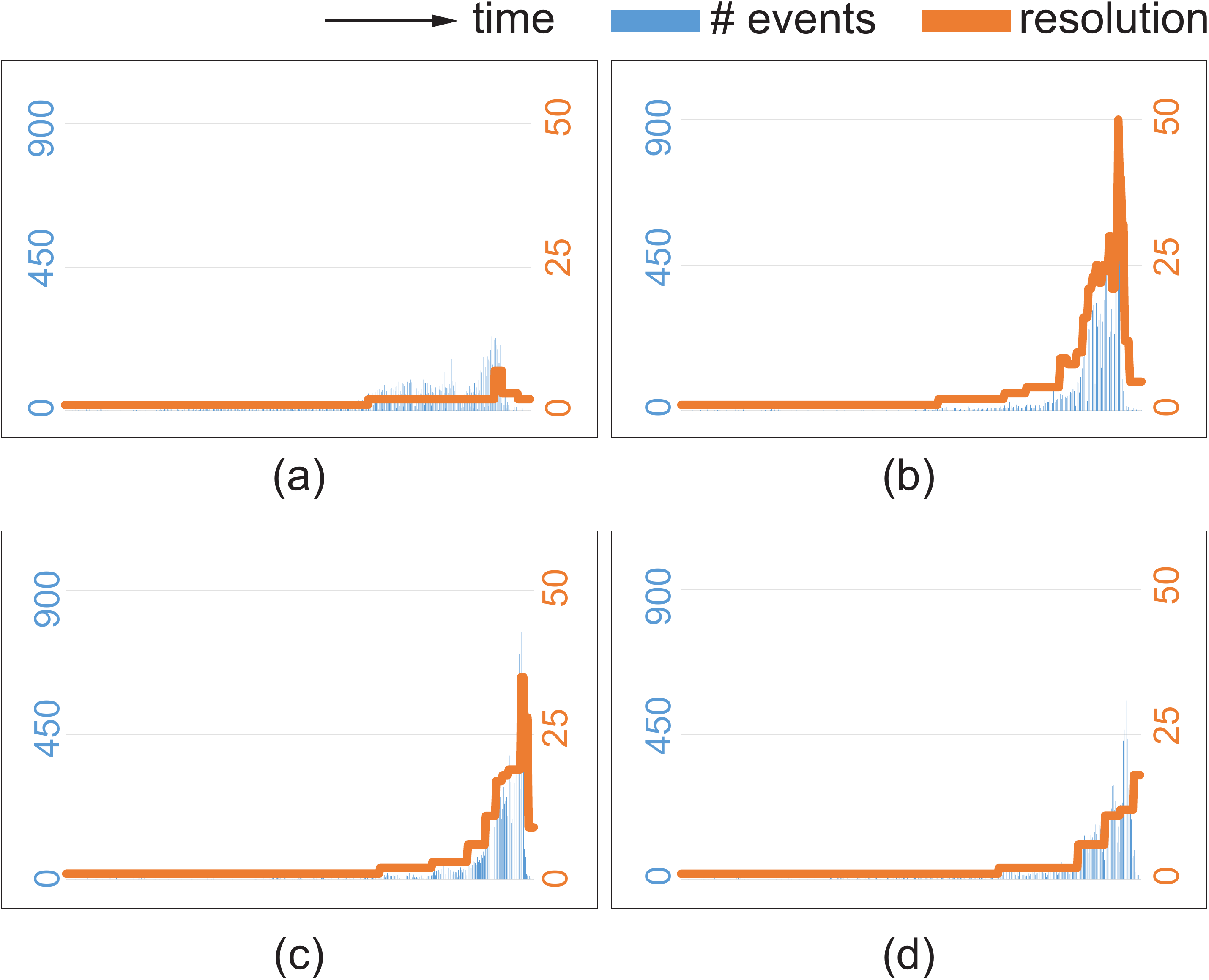}
	\centering
	\caption{Our nonuniform timeslicing and the relation between the adopted resolution scales and the event distribution for the Enron network. (a) $w_{size} = 100$ and $FF=0.9$ (921 timestamps). (b) $w_{size} = 50$ and $FF=0.99$ (357 timestamps). (c) $w_{size} = 100$ and $FF=0.99$ (448 timestamps). (d) $w_{size} = 200$ and $FF=0.99$ (579 timestamps).}
	%Plots (a,c): different Fading Factor values ($FF$) for the same window size ($w_{size} = 100$); Plots (b,c,d): Different windows sizes for $FF = 0.99$.}
	\label{fig:graficosEnron}
\end{figure}

Figure~\ref{fig:graficosEnron} presents our method's behaviour under different values of $w_{size}$ and $FF$ for the Enron network. Comparing the plots (a,c), it is possible to see the impact of the fading factor in the resolution computation. As can be seen, the high number of events near the end of the network is reflected in the timeslicing for the two $FF$ values tested. Comparing the plots (b,c,d), one can see how frequent the timeslicing occurs according to the window size. As discussed, large windows make the change in the resolution scale less frequent and, as a consequence, each resolution may not faithfully represent the different number of events and their distribution. One can see such situation occurring in the Enron network by analysing the resolution evolution under $w_{size} = 200$ and $FF = 0.99$ (Fig.~\ref{fig:graficosEnron}(d)): at the end of the network, the number of events decreases abruptly, but the resolution scale remains high.

\begin{figure*}[ht]
	\includegraphics[width=0.85\linewidth]{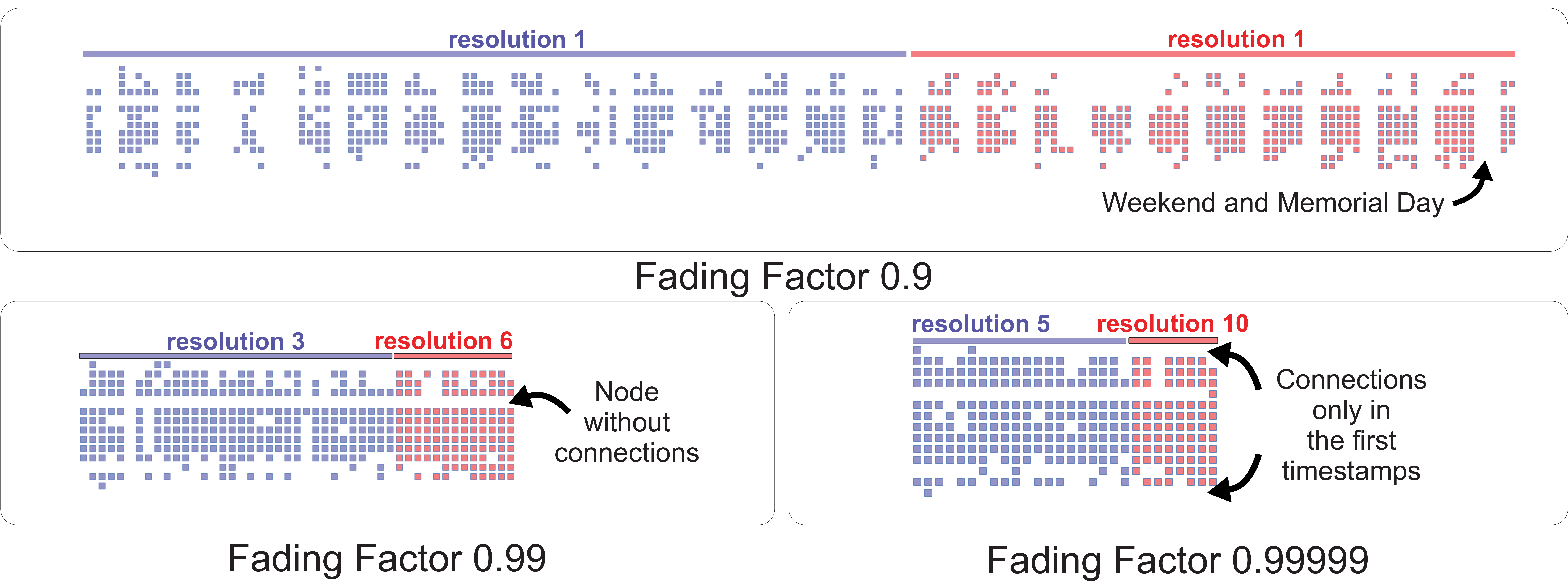}
	\centering
	\caption{Impact of different Fading Factor ($FF$) values on the layout ($w_{size} = 100$). Different FF values lead to different visual patterns. The change in the node colour represents a change in the resolution scale (new timeslice). Node ordering defined by Recurrent Neighbors~\cite{DyNetVis} using resolution~1.}
	%\caption{Impact of different Fading Factor ($FF$) values on the layout ($w_{size} = 100$). The change of the node color represents a change in the resolution value. In the layout using $FF = 0.9$, weekends and holidays are easily identified. In the layouts with $FF = 0.99$ and $FF = 0.99999$, it is easy to find the node without any connections and the presence of the first and the last nodes only in the first timestamps.}
	\label{fig:imgFds}
\end{figure*}

Figure~\ref{fig:imgFds} shows an approximation of the same time interval (near Dec. $12^{th}$, 1999 to near May $31^{th}$, 2000)\footnote{Since the resolution scale may aggregate different days in a single timestamp, the first timestamp may also represent few days before the first day of the interval depending on the adopted resolution. In the same way, the last timestamp may also represent few days after the last day of the interval.} and the same group of nodes in three distinct layouts obtained by adopting $w_{size} = 100$ and three different Fading Factor values ($FF = 0.9$, $FF = 0.99$ and $FF = 0.99999$). The first layout, with $FF = 0.9$, maintained the original resolution scale during the whole interval. By doing so, each timestamp refers to a 1-day interval and so it was possible to identify days without events. Such days are usually weekends and holidays, such as the highlighted weekend May $28^{th}-29^{th}$, 2000 and holiday May $30^{th}$, 2000 (Memorial Day). By adopting $FF = 0.99$, one can see that the weekend/holiday pattern is lost due to the aggregation of days in a single timestamp. Another pattern, however, is revealed: it is easier to identify a node without interactions, i.e., a person in the company that did not receive or send any emails in this period. By analysing the layout with $FF = 0.99999$, it is possible to see that the node without interactions from the previous layout appears in the network in the last timestamp. Moreover, one can notice that the first and the last nodes of the layout had interactions only in the first timestamps. These last two patterns are visible in all three layouts, but they are more easily perceived in the layouts with higher $FF$ values.

\begin{figure}[!ht]
	\includegraphics[width=0.8\linewidth]{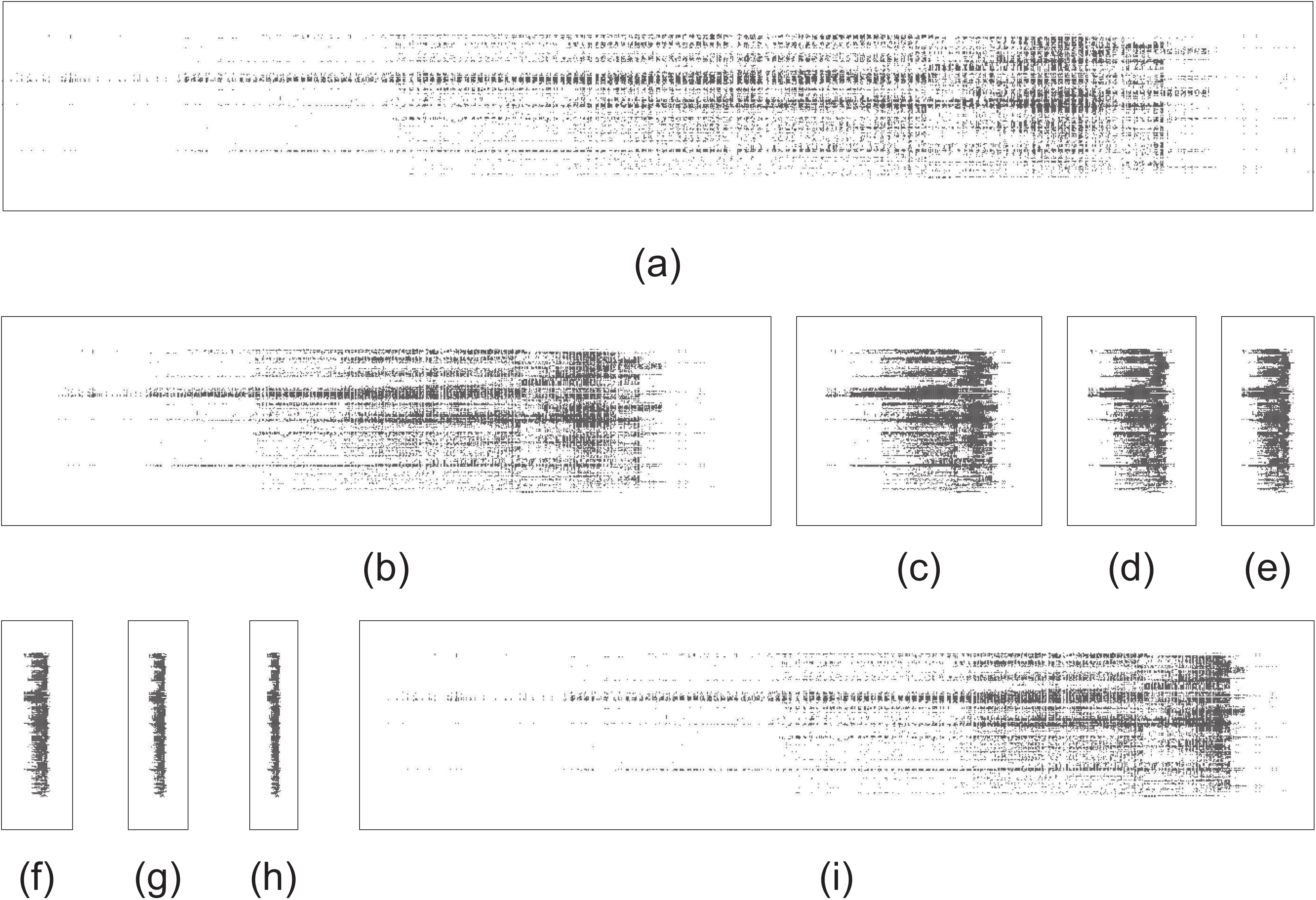}
	\centering
	\caption{TAM layouts for the Enron network considering different resolution scales. (a) Res.~1. (b) Res.~2. (c) Res.~7. (d) Res.~15. (e) Res.~25. (f) Res.~50. (g) Res.~75. (h) Res.~100. (i) Our method ($w_{size} = 100$ and $FF = 0.9$). Node ordering defined by Recurrent Neighbors~\cite{DyNetVis} using Res~1.}
	\label{fig:full_enron}
\end{figure}

Figure~\ref{fig:full_enron} shows TAM layouts for the Enron network considering different timeslicing approaches. Figure~\ref{fig:full_enron}(a-h) shows the TAM layouts for uniform timeslicing using resolutions 1, 2, 7, 15, 25, 50, 75, and 100, respectively. Figure~\ref{fig:full_enron}(i) shows the TAM layout generated by our method ($w_{size} = 100$ and $FF = 0.9$). Resolutions 1, 2, and 7 (Fig.~\ref{fig:full_enron}(a-c)) are shown because they represent the lower (and original), the average and the higher resolution scales adopted by our method for this network. The other resolutions (Fig.~\ref{fig:full_enron}(c-h)) are arbitrary values.

\begin{figure*}[ht]
	\includegraphics[width=\linewidth]{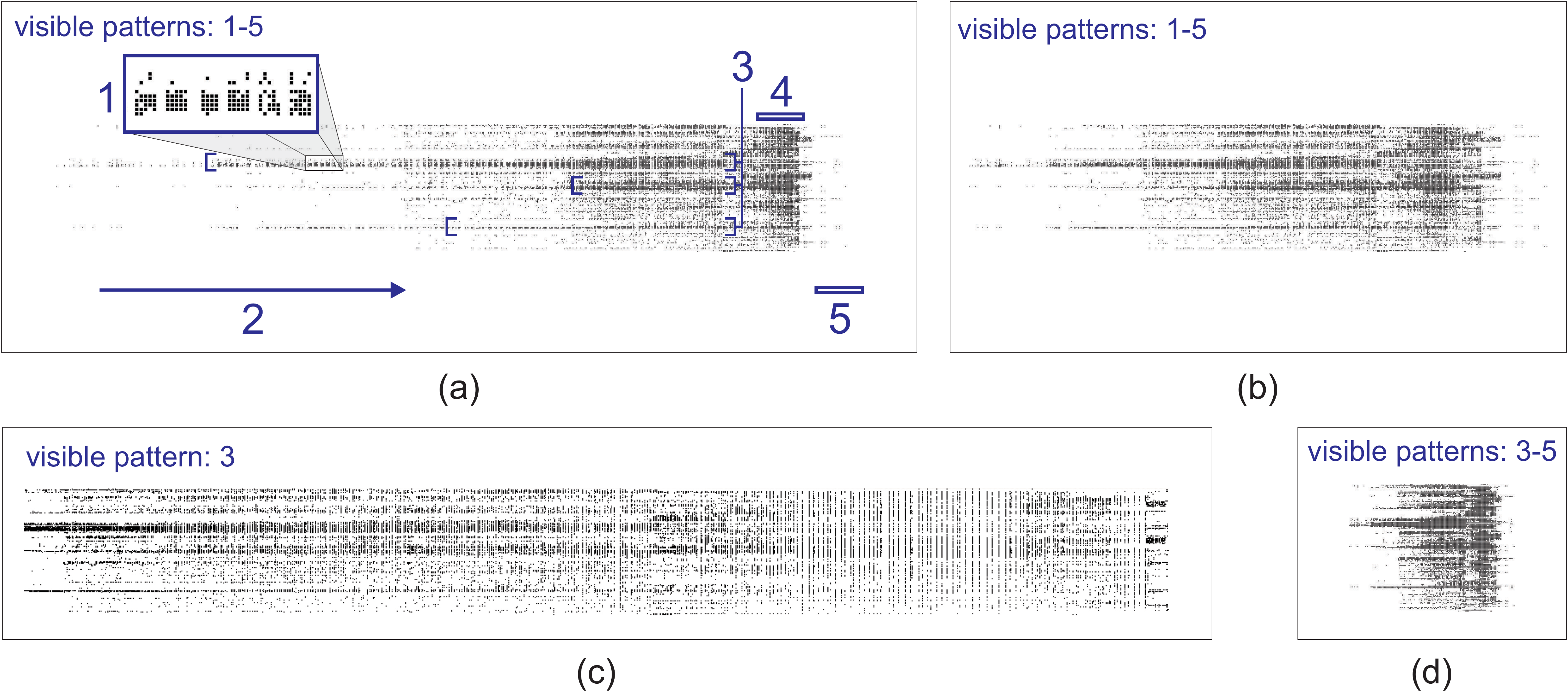}
	\centering
	\caption{TAM layouts generated by different timeslicing approaches and their visible patterns in the Enron network. (a) Our method (921 timestamps, $w_{size} = 100$ and $FF = 0.9$). (b) Resolution 2 (673 timestamps). (c) BVC (1,345 timestamps). (d) Resolution 7 (193 timestamps). Depending on the layout, a maximum of five visual patterns can be identified: (1) weekdays, in which there are interactions among nodes, and weekends, that are days without interactions; (2) perception of the growing number of events over time; (3) identification of highly active groups of nodes; (4) a time interval with a burst of events near the end of the network; and (5) abrupt decrease in the number of events followed by the end of the network. Node ordering defined by Recurrent Neighbors~\cite{DyNetVis} using resolution~1.}
	\label{fig:comparaPadroesEnron}
\end{figure*}

As illustrated in Fig.~\ref{fig:comparaPadroesEnron}, depending on the timeslicing being used, more or less patterns can be identified. The layout generated by our method allows the identification of at least 5 patterns (Fig.~\ref{fig:comparaPadroesEnron}(a)): (1) weekdays, in which there are interactions among nodes, and weekends (without interactions); (2) perception of the growing number of events over time; (3) identification of highly active groups of nodes; (4) a time interval with a burst of events near the end of the network; and (5) abrupt decrease in the number of events followed by the end of the network. The uniform timeslicing using resolution~2 (Fig.~\ref{fig:comparaPadroesEnron}(b)) also allows the perception of all five patterns. However, recall that this resolution represents the average value adopted by our method, which supports the claim that it chooses resolution scales that are indeed suitable for the network analysis. As expected, BVC redistributed the events along the timestamps, and so these temporal patterns (all but pattern~3) are lost (Fig.~\ref{fig:comparaPadroesEnron}(c)). By using resolution 7 in a uniform timeslicing (Fig.~\ref{fig:comparaPadroesEnron}(d)), patterns~1 and~2 are lost. Each timestamp in this resolution represents 7 days and so there is no separation of weekdays and weekends or the perception of growing node activity. One can note that layouts with temporal resolutions above 7 (see Fig.~\ref{fig:full_enron}(d-h)) are even worse for the Enron network visual analysis.

The ideal timeslicing depends on the network being analysed. For the Primary School network, the uniform timeslicing using resolution~25 allowed the identification of several patterns (see Fig.~\ref{fig:comparaPadroesEscola}(b)). On the other hand, resolution 25 is not a good choice for the Enron network (Fig.~\ref{fig:full_enron}(d)). In the same way, a uniform timeslicing using resolution~2 would not improve Primary School analysis. Our method is capable of adapting the resolution scale according to the number and distribution of events, thus enhancing the network visual analysis.

\begin{figure*}[h]
	\includegraphics[width=\linewidth]{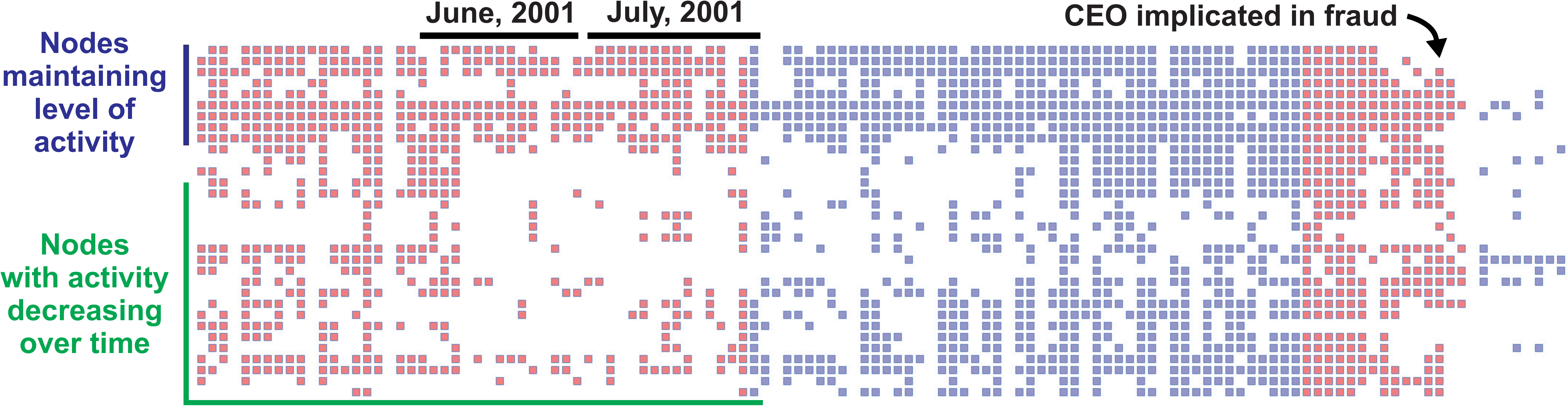}
	\centering
	\caption{TAM layout generated by our method ($w_{size} = 100$ and $FF = 0.9$) showing a portion of the network. Two patterns are visible: (i) a decrease in the number of events in June and July 2001; and (ii) an abrupt decrease in the number of events followed by the end of the network. The change in the node colour represents a change in the resolution scale (new timeslice). Node ordering defined by Recurrent Neighbors~\cite{DyNetVis} using resolution~1.}
	%\caption{Two patterns observed in a TAM layout with the adaptive resolution ($w_{size} = 100$ and $FF = 0.9$). In the first one, it is possible to see a decrease in the number of connections in June and July, 2001 (possibly related to the event ``Rove divests his stocks in energy''). In the second one, there is an abrupt decrease in the number of connections followed by the end of the network (related to the event ``Lay [Enron CEO] implicated in plot to inflate profits and hide losses''). The change of the node colors represents a change in the resolution value. Node ordering defined by Recurrent Neighbors~\cite{DyNetVis}.}
	\label{fig:doisemum}
\end{figure*}

\begin{figure}[ht]
	\includegraphics[width=0.6\linewidth]{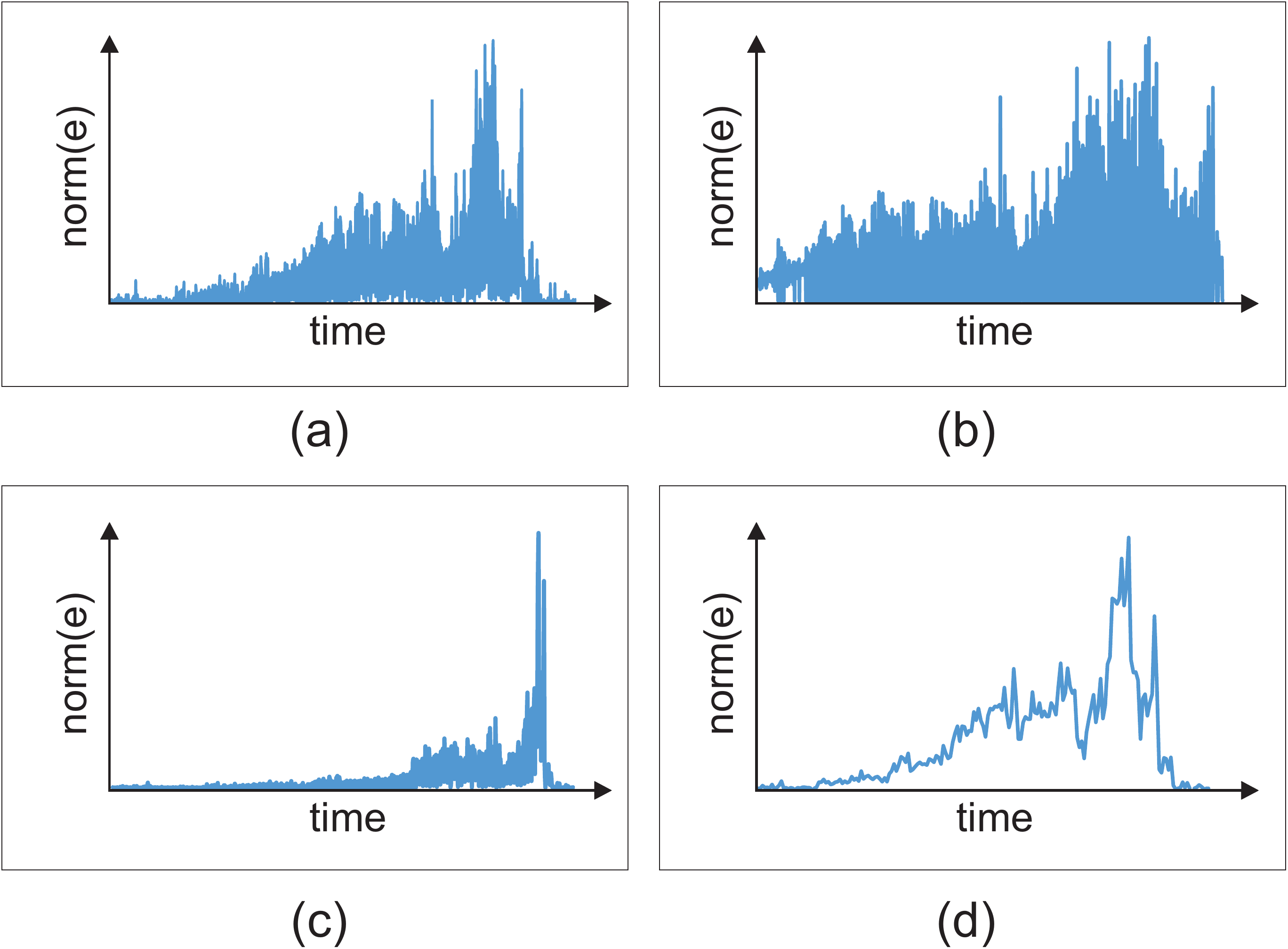}
	\centering
	\caption{Spread of events over time according to different timeslicing approaches for the Enron network. (a) Original resolution (1,346 timestamps). (b) BVC (1,345 timestamps). (c) Our method (921 timestamps, $w_{size} = 100$ and $FF = 0.9$). (d) Res.~25 (193 timestamps). While BVC changes the event distribution because of its histogram equalisation procedure, our method provides a distribution similar to those from uniform approaches. ``norm(e)'' refers to the normalisation of the number of events to values between~0 and~1.}
	\label{fig:graficosEnronJeitao1}
\end{figure}

Figure~\ref{fig:doisemum} presents two other patterns observed in the layout generated by our method when \textit{zooming in} the time interval with a burst of events showed in Fig.~\ref{fig:comparaPadroesEnron}(a). According to Sun et al.~\cite{graphScope}, in June 2001 occurred an important company episode related to the Enron accounting fraud: \textit{``Rove divests his stocks in energy''}. In the layout, it is possible to see a decrease in the number of events (emails) in the majority of the days in June and July involving the majority of the nodes. Such pattern may be related to this important episode. The layout also shows the moment in which there is an abrupt decrease in the number of events followed by the end of the network. Such decrease is related to another company episode, \textit{``Lay [Enron CEO] implicated in plot to inflate profits and hide losses''}~\cite{graphScope}, which happened in Feb $4^{th}$, 2002. After the decrease of events, our method changed the resolution scale from 7 to 3, reflecting the new number of events. Temporal patterns such as these are probably lost when using BVC because of its event redistribution (Fig.~\ref{fig:graficosEnronJeitao1}(b)). Our method (Fig.~\ref{fig:graficosEnronJeitao1}(c)), on the other hand, provides a distribution similar to those from uniform approaches (Fig.~\ref{fig:graficosEnronJeitao1}(a,d)), and thus is capable of highlighting such temporal patterns.

\section{Limitations}
\label{limitations}

When timeslicing, one should be aware that events may be lost to improve network comprehension (due to the consecutive timestamp grouping), and so relevant information may be lost in the process. Such characteristic exists in any other sampling strategy. Our method, however, considers the number of events and maintains their non-stationary distribution in an attempt to reduce such impairment.

Although our method improves the layout by manipulating the network temporal dimension, the node positioning represents another important aspect that has to be considered, since the ordering quality may impact layout readability. We thus recommend the adoption of a high-quality node ordering method. Eventually, the joint employment of sampling strategies may be required.

Since two timestamps in the layout may represent completely different time intervals, one should pay attention in the resolution scale adopted in each of them when the task depends on this information (e.g., to decide which node has been active for the longest time in the network). Changes in the node colour, as used in the experiments, attenuate this limitation, but other visual encoding can be used. In such cases, where the nonuniform timeslicing impairs the analysis, our method remains useful as the average resolution scale computed by it represents a good choice for a uniform timeslicing (as occurred with resolution 25 in our Primary School analysis).

Finally, we have demonstrated our method's quality using TAM and MSV. Although our method runs online, these visual representations draw the network elements (nodes and/or edges) in an offline manner. This is a characteristic of these layouts and not a limitation of our method. Although they could be adapted to handle online scenarios by plotting consecutive windows over time, this adaptation is out of the scope of this paper. Furthermore, our method does not rely on particular layouts' characteristics (e.g., length/positioning of edges or animated vs timeline layouts) and thus could be applied in different layouts as well. In animated layouts, however, the visual analysis would probably be impaired in some cases, since the number of frames devoted to high-activity periods would reduce, potentially breaking the user's mental map.

\section{Conclusion}
\label{conclusion}

We proposed in this paper an online and nonuniform timeslicing method for network visualisation that highlights temporal patterns such as bursts of events, highly-active groups of nodes, and others. Without it, when handling temporal networks one should test different uniform timeslices until the ``less worst'' is found. Besides the effort of such preliminary tests, analyses of different networks require different temporal resolutions. For streaming scenarios, exploratory analysis to support the timeslicing may not be possible because events are arriving online and in non-stationary distribution. For the same reason, considering an initial set of events to support the choice of the resolution scale may be inefficient as well.

Our method considers the number of events and their distribution to adapt the layout. This is possible because the choice of each new resolution scale uses only events from a sliding window, with old information being discounted by a forgetting mechanism. The method has low time and spatial computational complexity, since there is no need for various scans in the data and edges can be discarded once they are processed. In our experiments, we have analysed two real-world networks with different characteristics and the results show that the resolution scales automatically adopted are indeed suitable for each network analysis.

As future work, we intend to perform user experiments to validate our method considering the quality of the produced layout. Besides, the choice of both window size and fading factor value directly affects the layout. These are currently user-dependent parameters and we will try to automate them. We also intend to apply our method in other network visualisation strategies.

%%
%% The acknowledgments section is defined using the "acks" environment
%% (and NOT an unnumbered section). This ensures the proper
%% identification of the section in the article metadata, and the
%% consistent spelling of the heading.
\section*{Acknowledgments}
This research was supported by Conselho Nacional de Desenvolvimento Cient\'ifico e Tecnol\'ogico - CNPq, Coordena\c{c}\~ao de Aperfei\c{c}oamento de Pessoal de N\'ivel Superior (CAPES PrInt - Grant number 88881.311513/2018-01), and Funda\c{c}\~ao de Amparo a Pesquisa do Estado de Minas Gerais - FAPEMIG.

\bibliographystyle{eg-alpha-doi} 
\bibliography{main}    

\end{document}